\documentclass[fdp,a4paper,fleqn%
]{w-art}
\usepackage{times,cite,w-thm}
\theoremstyle{plain}

\theoremstyle{definition}

\usepackage[]{graphicx}
\begin{document}
%%    The information for the title page will be placed between
%%    \begin{document} and \maketitle. The order of most entries
%%    is determined by the class file and can not be changed by
%%    rearranging them. The maketitle command follows after the
%%    abstract.
%%
%%    Most of the following commands will be completed by the publisher.
%%
%%    The copyrightyear is defined in the .clo file as the first argument
%%    of the copyrightinfo command. If the copyrightyear differs from that
%%    value it might be adjusted by the following definition:
%%
%% \renewcommand{\copyrightyear}{2007}% uncomment to change the copyrightyear.
%%
\DOIsuffix{theDOIsuffix}
%%
%% issueinfo for the header line
\Volume{55}
\Month{01}
\Year{2007}
%%
%%    First and last pagenumber of the article. If the option
%%    'autolastpage' is set (default) the second argument may be left empty.
\pagespan{1}{}
%%
%%    Dates will be filled in by the publisher. The 'reviseddate' and
%%    'dateposted' (Published online) entry may be left empty.
%\Receiveddate{XXXX}
%\Reviseddate{XXXX}
%\Accepteddate{XXXX}
%\Dateposted{XXXX}
%%
%\keywords{Black holes, entropy, heterotic string theory.}
%\subjclass[pacs]{04.50.Gh, 11.25.-w,11.30.Pb, 04.70.Dy}
%  (Please use PACS-codes from the enclosed list
 % (ASCII2006FullPACS.txt) or from www.aip.org/pacs)

%% \pretitle{Editor's Choice}

%% We have a short and a long form for the title. The short form
%% (optional argument) goes into the running head.

\title[Black holes in heterotic string theory]{Five-dimensional black holes
       in heterotic string theory\footnote{Based on talks presented by P.D.P. 
       at the III Southeastern European Workshop \emph{Challenges Beyond the 
       Standard Model} (September 2-9, 2007, Kladovo, Serbia), and 
       DFG \& NZZ Workshop on Field Theory, Non-commutative Geometry
       and Strings (November 9-11, 2007, Zagreb, Croatia)}}

%% Please do not enter footnotes or \inst{}-notes into the optional
%% argument of the author command. The optional argument will go into
%% the header.  If there is only one address the marker \inst{x} may be
%% omitted.

%% Information for the first author.
\author[M. Cvitan]{Maro Cvitan\inst{1,3}%
%  \footnote{E-mail:~\textsf{cvitan@sissa.it}}
}
\address[\inst{1}]{International School for Advanced Studies (SISSA/ISAS),
        Via Beirut 2--4, 34014 Trieste, Italy}
%%
%%    Information for the second author
\author[P. Dominis Prester]{Predrag Dominis Prester\inst{2,3}%
  }
\address[\inst{2}]{Physics Department, Faculty of Arts and Sciences,
        University of Rijeka\\ Omladinska 14, HR-51000 Rijeka, Croatia}
%%
%%    Information for the third author
\author[A. Ficnar]{Andrej Ficnar\inst{3}%
%  \footnote{E-mail:~\textsf{aficnar@fizika.org}.}
}
\address[\inst{3}]{Theoretical Physics Department, Faculty of Science,
        University of Zagreb\\ p.p. 331, HR-10002 Zagreb, Croatia}
\author[S. Pallua]{Silvio Pallua\inst{3}%
%  \footnote{E-mail:~\textsf{pallua@phy.hr}.}
}
\author[I. Smoli\'{c}]{Ivica Smoli\'{c}\inst{3}%
%  \footnote{E-mail:~\textsf{ismolic@phy.hr}.}
}
%%
%%    \dedicatory{This is a dedicatory.}
\begin{abstract}
We review recent results on near-horizon static black hole solutions
and entropy in $R^2$-corrected $N=2$ SUGRA in $D=5$, focusing on 
actions connected to heterotic string compactified on $K3\times S^1$. 
Comparison with $\alpha'$-perturbative results, results obtained by 
using simple Gauss-Bonnet $R^2$-correction, OSV conjecture and 
microscopic stringy description (for small black holes) shows that
situation in $D=5$ is, in a sense, even more interesting then in 
$D=4$.
\end{abstract}
%% maketitle must follow the abstract.
\maketitle                   % Produces the title.

%% If there is not enough space inside the running head
%% for all authors including the title you may provide
%% the leftmark in one of the following three forms:

%% \renewcommand{\leftmark}
%% {First Author: A Short Title}

%% \renewcommand{\leftmark}
%% {First Author and Second Author: A Short Title}

%% \renewcommand{\leftmark}
%% {First Author et al.: A Short Title}

%% \tableofcontents  % Produces the table of contents.

%\section{Introduction}

\section{5-dimensional black holes in higher derivative $N=2$ SUGRA}

Bosonic part of the Lagrangian for the $N=2$ supergravity action in 
five dimensions is given by
\begin{eqnarray} \label{l0susy}
4\pi^2\mathcal{L}_0 &=& 2 \partial^a \mathcal{A}^\alpha_i \partial_a
\mathcal{A}_\alpha^i + \mathcal{A}^2 
\left(\frac{D}{4}-\frac{3}{8}R-\frac{v^2}{2}\right)
+ \mathcal{N} \left(\frac{D}{2}+\frac{R}{4}+3v^2\right)
+ 2 \mathcal{N}_I v^{ab} F_{ab}^I \nonumber \\ 
&& + \mathcal{N}_{IJ} \left( \frac{1}{4} F_{ab}^I F^{Jab} 
 + \frac{1}{2} \partial_a M^I \partial^a M^J \right)
+ \frac{e^{-1}}{24} c_{IJK} A_a^I F_{bc}^J F_{de}^K \epsilon^{abcde}
\end{eqnarray}
where $\mathcal{A}^2 = \mathcal{A}^\alpha_{i\,ab}
\mathcal{A}_\alpha^{i\,ab}$ and $v^2 = v_{ab}v^{ab}$. Also,
\begin{equation}
\mathcal{N} = \frac{1}{6} c_{IJK} M^I M^J M^K , \quad
\mathcal{N}_I = \partial_I \mathcal{N} = \frac{1}{2} c_{IJK} M^J M^K
, \quad 
\mathcal{N}_{IJ} = \partial_I \partial_J \mathcal{N} = c_{IJK} M^K
\end{equation}

A bosonic field content of the theory is the following. We have Weyl
multiplet which contains the f\"{u}nfbein $e_\mu^a$, the two-form 
auxiliary field $v_{ab}$, and the scalar auxiliary field $D$. There
are $n_V$ vector multiplets enumerated by $I=1,\ldots,n_V$, each
containing the one-form gauge field $A^I$ (with the two-form field 
strength $F^I=dA^I$), and the scalar $M^I$. Scalar fields 
$\mathcal{A}_\alpha^i$, which are belonging to the hypermultiplet, can
be gauge fixed and the convenient choice is given by
%\begin{equation} \label{hgfix}
$\mathcal{A}^2 = -2$, $\partial_a \mathcal{A}^\alpha_i = 0$.
%\end{equation}

Lagrangian (\ref{l0susy}) can be obtained from 11-dimensional SUGRA by
compactifying on six-dimensional Calabi-Yau spaces. Then $M^I$
have interpretation as moduli (volumes of $(1,1)$-cycles), and $c_{IJK}$
as intersection numbers. Condition $\mathcal{N}=1$ is a condition of real
special geometry.

Action (\ref{l0susy}) is invariant under SUSY variations, which
when acting on the purely bosonic configurations are given with
\begin{eqnarray} \label{svar}
\delta\psi_\mu^i &=& \mathcal{D}_\mu\varepsilon^i + \frac{1}{2}v^{ab}
 \gamma_{\mu ab}\varepsilon^i - \gamma_\mu\eta^i \nonumber \\
\delta\xi^i &=& D\varepsilon^i 
 - 2\gamma^c\gamma^{ab}\varepsilon^i\mathcal{D}_a v_{bc}
 - 2\gamma^a\varepsilon^i\epsilon_{abcde}v^{bc}v^{de} 
 + 4\gamma\cdot v\eta^i \nonumber \\
\delta\Omega^{Ii} &=& - \frac{1}{4}\gamma\cdot F^{I}\varepsilon^i
 - \frac{1}{2}\gamma^a\partial_a M^{I}\varepsilon^i - M^{I}\eta^i
 \nonumber \\
\delta\zeta^{\alpha} &=& 
 \left(3\eta^j-\gamma\cdot v\varepsilon^j\right)\mathcal{A}_j^\alpha
\end{eqnarray}
where $\psi_\mu^i$ is gravitino, $\xi^i$ auxiliary Majorana spinor
(Weyl multiplet), $\delta\Omega^{Ii}$ gaugino (vector multiplets), and
$\zeta^{\alpha}$ is a fermion field from hypermultiplet.

In \cite{Hanaki:2006pj} a four-derivative part of the action was
constructed by supersymmetric completion of the mixed
gauge-gravitational Chern-Simons term 
$A \land \textrm{tr} (R \land R)$. The bosonic part of the action is
\begin{eqnarray} \label{l1susy}
4\pi^2\mathcal{L}_1 &=& \frac{c_{I}}{24} \left\{ \frac{e^{-1}}{16}
\epsilon_{abcde} A^{Ia} C^{bcfg} C^{de}_{\;\;\;\,fg} 
+ M^I \left[ \frac{1}{8} C^{abcd} C_{abcd} + \frac{1}{12} D^2 
 - \frac{1}{3} C_{abcd} v^{ab} v^{cd} 
\right. \right. \nonumber \\ &&
 + 4 v_{ab}v^{bc} v_{cd} v^{da} - (v_{ab}v^{ab})^2
 + \frac{8}{3} v_{ab} \hat{\mathcal{D}}^b \hat{\mathcal{D}}_c v^{ac}
 + \frac{4}{3} \hat{\mathcal{D}}^a v^{bc} \hat{\mathcal{D}}_a v_{bc}
 + \frac{4}{3} \hat{\mathcal{D}}^a v^{bc} \hat{\mathcal{D}}_b v_{ca}
\nonumber \\ && \left. 
 - \frac{2}{3} e^{-1} \epsilon_{abcde} v^{ab} v^{cd}
   \hat{\mathcal{D}}_f v^{ef} \right] 
+ F^{Iab} \left[ \frac{1}{6} v_{ab} D - \frac{1}{2} C_{abcd} v^{cd}
 + \frac{2}{3} e^{-1} \epsilon_{abcde} v^{cd} 
   \hat{\mathcal{D}}_f v^{ef} 
\right. \nonumber \\ && \left. \left.
 + e^{-1} \epsilon_{abcde} v^{c}_{\;f} \hat{\mathcal{D}}^d v^{ef}
 - \frac{4}{3} v_{ac}v^{cd} v_{db} - \frac{1}{3} v_{ab} v^2 \right]
\right\}
\end{eqnarray}
where $c_I$ are some constant coefficients\footnote{From the viewpoint of
compactification of $D=11$ SUGRA they are topological numbers connected
to second Chern class.}, $C_{abcd}$ is the Weyl 
tensor, and $\hat{\mathcal{D}}_a$ is the conformal covariant derivative.

We are interested in extremal black hole solutions of the action
obtained by combining (\ref{l0susy}) and (\ref{l1susy}):\footnote{Our
conventions for Newton coupling is $G_5=\pi^2/4$ and for the string
tension $\alpha'=1$.}
\begin{equation} \label{lsusy}
\mathcal{A} = \int dx^5 \sqrt{-g} \mathcal{L} 
 = \int dx^5 \sqrt{-g} (\mathcal{L}_0 + \mathcal{L}_1)
\end{equation}

The action (\ref{lsusy}) is quartic in derivatives and generally
probably too complicated for finding complete analytical black hole
solutions even in the simplest spherically symmetric case. But, if one
is more modest and interested just in a near-horizon behavior (which is
enough to find the entropy) of {\em extremal} black holes, there is a
smart way to do the job - Sen's entropy function formalism
\cite{Sen:2005wa}.

For five-dimensional spherically symmetric extremal black holes 
near-horizon geometry is expected to be $AdS_2\times S^3$, which has 
$SO(2,1)\times SO(4)$ symmetry. If the Lagrangian can be written in a 
manifestly diffeomorphism covariant and gauge invariant way, it is 
expected that near the horizon the complete background should respect 
this symmetry. In our case it means that near-horizon geometry should
be given with
\begin{eqnarray} \label{efhere}
&& ds^2 = v_1 \left( -x^2 dt^2 + \frac{dx^2}{x^2} \right)
 + v_2\,d\Omega_{3}^2 \nonumber \\
&& F^{I}_{tr}(x) = -e^I \;,\qquad v_{tr}(x) = V \;,\qquad
% \nonumber \\ && 
 M^I(x) = M^I \;, \qquad D(x) = D
\end{eqnarray}
where $v_i$, $e^I$, $M^I$, $V$, and $D$ are constants. All covariant 
derivatives are vanishing. If one defines
\begin{equation}\label{fdef}
f = \int_{S^3} \sqrt{-g} \, \mathcal{L} \;,
\end{equation}
where right hand side is evaluated on the background (\ref{efhere}),
then equations of motion are equivalent to
\begin{equation} \label{seom}
0 = \frac{\partial f}{\partial v_1} \;, \qquad
0 = \frac{\partial f}{\partial v_2} \;, \qquad
0 = \frac{\partial f}{\partial M^I} \;, \qquad
0 = \frac{\partial f}{\partial V} \;, \qquad
0 = \frac{\partial f}{\partial D} \;.
\end{equation}
Derivatives over electric field strengths $e^I$ are giving (properly 
normalized) electric charges:
\begin{equation}\label{chgdef}
q_I = \frac{\partial f}{\partial e^I}
\end{equation}
The entropy (equal to the Wald formula \cite{Wald}) is given with
\begin{equation} \label{entropy}
S_{bh} = 2\pi \left( q_I \, e^I - f \right)
\end{equation}

It is immediately obvious that though the system (\ref{seom}), (\ref{chgdef})
is algebraic, it is in generic case too complicated to be solved in direct 
manner, and that one should try to find some additional information. Such 
additional information can be obtained from supersymmetry. It is
known that there should be 1/2 BPS black hole solutions, for which it
was shown in \cite{Chamseddine:1996pi} that near the horizon
supersymmetry is enhanced fully. This means that in this case we can
put all variations in (\ref{svar}) to zero, which one can use to
express all unknowns in terms of one. To fix remaining unknown we just
need one equation from (\ref{seom}), where the simplest is the one for
$D$. 

Typically one is interested in expressing the results in terms of
charges, not field strengths, and this is achieved by using
(\ref{chgdef}). One gets
\cite{Castro:2007hc,Alishahiha:2007nn,Cvitan:2007en}
\begin{equation} \label{mbareq}
8\,c_{IJK} \bar{M}^J \bar{M}^K = q_I + \frac{c_{I}}{8} \;,\qquad\qquad
\bar{M}^I \equiv \sqrt{v_1} M^I \;,
\end{equation}
where we introduced scaled moduli $\bar{M}^I$. The entropy becomes
\begin{equation} \label{sentm}
S_{bh}^{(BPS)} = \frac{8\pi}{3} c_{IJK}\bar{M}^I \bar{M}^J \bar{M}^K 
\end{equation}
A virtue of this presentation is that if one is interested only in
entropy, then it is enough to consider just (\ref{mbareq}) and
(\ref{sentm}), in which the sole effect of the higher derivative terms 
are just constant shifts of charges 
$q_I \to q_I+c_I/8$.\footnote{Those who are interested in full 
solutions can consult 
\cite{Castro:2007hc,Alishahiha:2007nn,Cvitan:2007en}.} It was 
shown in \cite{Castro:2007ci} that (\ref{sentm}) agrees with the OSV 
conjecture \cite{topstr}, after proper treatment of uplift from $D=4$ 
to $D=5$ is made.

\section{Heterotic black holes -- non-BPS solutions}

We shall be especially interested in the case when prepotential is
of the form
\begin{equation}
\mathcal{N} = \frac{1}{2}M^1c_{ij}M^iM^j \;, \qquad i,j>1
\end{equation}
where $c_{ij}$ is a regular matrix with an inverse $c^{ij}$.
In this case, which corresponds to $K3\times T^2$ 11-dimensional 
compactifications, it is easy to show that the entropy of BPS black 
holes is given with
\begin{equation} \label{seK3}
S_{bh}^{(BPS)}
 = 2\pi\sqrt{\frac{1}{2} \hat{q}_1 c^{ij}\hat{q}_i\hat{q}_j} \;,
\qquad \hat{q_I}=q_I+\frac{c_I}{8}
\end{equation}
When additionally $c_1=24$, $c_i=0$, our action is equivalent to the 
(consistently truncated) tree-level effective action of heterotic string 
compactified on $K_3 \times S^1$. 

One especially interesting (and simple) case is given with 
$STU$-prepotential $\mathcal{N} = M^1 M^2 M^3$, which corresponds to
heterotic string on $T^4 \times S^1$. Black hole solutions are
caracterised by three integer charges usually denoted as $m=q_1$,
$n=q_2$ and $w=q_3$.\footnote{In heterotic string language $n$
and $w$ are momentum and winding number on $S^1$, and $m$ is the
magnetic charge of antisymmetric $B_{\mu\nu}$ field.} Constructed BPS
black hole solutions are physically acceptable for $m\ge0$ and 
$n,w>0$. The entropy is now
\begin{equation} \label{Ssusyb}
S_{bh}^{(BPS)} = 2\pi\sqrt{nw(m+3)} \;.
\end{equation}
It is remarkable that here non-BPS solutions (for almost all values of 
charges) were also analytically constructed \cite{Cvitan:2007en}. For 
example, for $m\ge1$, $n<0$, $w>0$ the entropy is
\begin{equation}
S_{bh}^{(n-BPS)} = 2\pi\sqrt{|n|w (m-1/3)} \;.
\end{equation}
Properties of these non-BPS solutions suggest possibility that they are 
descending from BPS states either in $D>5$ and/or $N>2$ 
\cite{Cvitan:2007en}.

Motivated by results from $D=4$, we have studied solutions when 
$R^2$-correction is given purely by Gauss-Bonnet density. We obtained
that the entropy is different from the one following from $R^2$ SUSY action
(\ref{seK3}). This mismatch was not present in $D=4$.

\section{Perturbative calculations in $\alpha'$}

In view of above results, it is interesting to perturbatively
calculate entropy of large 5-dimensional 3-charge extremal black holes
up to $\alpha'^2$-order using low energy effective action of heterotic 
string (which is fully known only up to $\alpha'^2$-order). The main 
virtue is that this is a straightforward calculation giving 
unambiguous results. Using entropy function formalism, and taking special
care for non manifestly covariant gravitational Chern-Simons term
(using technique from \cite{Sahoo:2006pm}), we obtained for the entropy
of BPS black holes
\cite{Cvitan:2007hu}
\begin{equation} \label{Spertb}
\mathcal{S}_{bh}^{(BPS)} =  2\pi \sqrt{nwm}
 \left( 1 + \frac{3}{2m} - \frac{9}{8m^2} + O\left(m^{-3}\right)
\right) ,
 \qquad n,w,m>0\,,
\end{equation}
which is in agreement with the supersymmetric result, i.e., with
(\ref{Ssusyb}) after being expanded in $1/m$.

For non-BPS black holes we obtained the entropy
\begin{equation} \label{Spertn}
\mathcal{S}_{bh}^{(n-BPS)} = 2\pi \sqrt{|n|wm}
 \left( 1 + \frac{1}{2m} - \frac{1}{8\,m^2} + O\left(m^{-3}\right)
 \right) , \qquad n<0, \; w,m>0,
\end{equation}
which disagrees with both SUSY (\ref{Ssusyb}) and Gauss-Bonnet 
results already at $\alpha'$-order. Instead, our result (\ref{Spertn})
suggests the following formula
\begin{equation} \label{Segzn}
\mathcal{S}_{bh}^{(n-BPS)} = 2\pi \sqrt{|n|w(m+1)}\, .
\end{equation}
Furthermore, if we take BPS formula (\ref{Ssusyb}) for granted, then
we have been able to show that $\alpha'^3$ term in the non-BPS entropy 
formula (\ref{Spertn}) must be $1/(16\,m^3)$, which is again in 
agreement with the conjectured expression (\ref{Segzn}). Now, using 
AdS/CFT arguments, from (\ref{Spertb}) and (\ref{Spertn}) one infers 
that central charges satisfy $c_L - c_R = 12w$, which is indeed what
is expected \cite{Kraus:2007vu}.

\section{Small black holes}

Extremely interesting is what happens when one takes $q_1=0$ in
(\ref{seK3}). For the $K3 \times S^1$ heterotic compactifications the 
entropy becomes
\begin{equation} \label{smallK3}
S_{bh}^{(BPS)}
 = 2\pi\sqrt{\frac{3}{2} c^{ij} q_i q_j} \;.
\end{equation}
On the other hand, for the action with Gauss-Bonnet $R^2$ term we 
obtain
\begin{equation} \label{smallGB}
S_{bh}^{(BPS)}
 = 4\pi\sqrt{\frac{1}{2} c^{ij} q_i q_j} \;.
\end{equation}
These black holes are \emph{small}, meaning that their horizon is generated
(regularized) by higher-derivative terms in the action.

Contrary to the large black holes discussed before, for these small 
black holes microscopic stringy description \emph{is} known (in the special
2-charge case of $T^4\times S^1$ heterotic compactification 
microstates are well-known perturbative Dabholkar-Harvey states) for
which statistical entropy was calculated 
\cite{Dabholkar:2004yr,Huang:2007sb}. For BPS states microscopic 
entropy (for $nw\gg1$) is \emph{exactly} equal to the black hole 
entropy obtained from the action supplemented with just Gauss-Bonnet 
$R^2$-correction (\ref{smallGB}), and disagrees with the entropy 
obtained by using supersymmetric $R^2$-correction 
(\ref{smallK3}).\footnote{For 2-charge black holes in $D>5$ one needs
in the action also higher Gauss-Bonnet densities 
\cite{Prester:2005qs}.}

\section{Conclusion}

We have analysed near-horizon solutions for (both BPS and non-BPS) 
static spherically symmetric black holes of $N=2$ SUGRA actions with
$R^2$-corrections, which are effective actions of tree-level 
heterotic string compactified on $K3\times S^1$ and $T^4\times S^1$.
In addition, we have also made calculations by taking for
$R^2$-correction just the Gauss-Bonnet density. In $D=5$, contrary to
a situation in $D=4$ (see \cite{Sen:2007qy} for a review), for these 
two types of higher-derivative corrections formulae for the entropy are
\emph{not} matching.  

For \emph{large} black holes, where full stringy microscopic 
description is not known, obtained entropies of BPS black holes are 
\emph{equal} to the one obtained from OSV conjecture and topological 
string, and are consistent with perturbative results up to 
$\alpha'^2$-order.

For \emph{small} black holes microscopic description is known, with
(asymptotic) statistical entropy \emph{exactly} matching the result 
obtained by using Gauss-Bonnet $R^2$-correction.

The exact matchings obtained by using just the $R^2$-corrections in 
effective actions (which has infinite expansion) is surprising. For
SUSY correction it has partial explanation through AdS$_3$/CFT$_2$ 
correspondence \cite{Kraus:2005vz}, but for small black holes where 
simple Gauss-Bonnet correction does the job (and SUSY correction 
fails) it is still a complete mistery. We believe that these issues 
deserve further investigation.

\begin{acknowledgement}
%  An acknowledgement may be placed at the end of the~article.
This work was supported by the Croatian Ministry of Science, Education
and Sport under the contract 119-0982930-1016. P.D.P. was also supported
by Alexander von Humboldt Foundation, and M.C. by Central European 
Initiative.
\end{acknowledgement}

% Use this code if you wish to generate your bibliography with BibTeX;
% please replace first the string "demo" below with the name(s) of
% the BibTeX data base(s) you want to use.
% The resulting bibliography-output (the contents of the .bbl file)
% must be pasted into this file before submission.
% 
% \bibliographystyle{pss}
% \bibliography{demo}
% 
% Replace the following example bibliography with your references
% before submission:

\end{document}